# Quantum decoherence of nitrogen-vacancy spin ensembles in a nitrogen spin bath in diamond under dynamical decoupling

Huijin Park[1,2,3], Mykyta Onizhuk[3], Eunsang Lee[4], Harim Lim[4], Junghyun Lee[3,4,5*], Sangwon Oh[2*], Giulia Galli[3,6,7], and Hosung Seo[1,2,4,8]*

[1]SKKU Advanced Institute of Nanotechnology (SAINT), Sungkyunkwan University, Suwon, 16419, Korea

[2]Department of Physics and Department of Energy Systems Research, Ajou University, Suwon, Gyeonggi 16499, Republic of Korea

[3]Pritzker School of Molecular Engineering, University of Chicago, Chicago, IL, USA

[4]Center for Quantum Information, Korea Institute of Science and Technology, Seoul 02792, Republic of Korea

[5]Division of Nano and Information Technology, KIST School, Korea University of Science and Technology, Seoul, Korea

[6]Department of Chemistry, University of Chicago, Chicago, IL, USA

[7]Center for Molecular Engineering and Materials Science Division, Argonne National Laboratory, IL, USA

[8]Department of Quantum Information Engineering, Sungkyunkwan University, Suwon, 16419, Korea

*Correspondence to: sangwonoh@ajou.ac.kr; jh_lee@kist.re.kr; seo.hosung@skku.edu

# ABSTRACT

The negatively charged nitrogen-vacancy (NV) center in diamond has emerged as a leading qubit platform for quantum technology applications. One of the key challenges for NV-based quantum applications is building an accurate model to predict its decoherence properties and their quantum nature. In this study, we combine theory and experiment to investigate NV decoherence dynamics in the presence of nitrogen donor (P1 center) baths, which is one of the dominant decoherence sources in diamond. We employ a cluster-correlation expansion (CCE) method to compute the NV decoherence under the Hahn-echo (HE) and Carr-Purcell-Meiboom-Gill (CPMG) pulse sequences at various P1 concentrations from 1ppm to 300 ppm. We show that the coherence time ($T_2$) increases with the number of $\pi$ pulses applied, indicating that the NV spin is decoupled from the P1 bath. Notably, we find that $T_2$ scales quadratically as a function of the pulse number, on a logarithmic scale, as opposed to the linear scaling predicted by widely accepted semi-classical theories in the literature. In our experiment, we measure the CPMG signal for two diamond samples with high P1 concentrations of 0.8ppm and 13ppm. We demonstrate that the $T_2$ scaling is indeed quadratic, thus confirming our theoretical predictions. Our results show that the quantum bath model combined with the CCE method can accurately capture the quantum nature of the P1-driven NV decoherence. Our study opens a new avenue for developing a complete noise model that could be used to optimize the performance of NV-based quantum devices.

# INTRODUCTION

The negatively charged nitrogen-vacancy (NV) center in diamond has been demonstrated to be a versatile qubit platform to enable robust and scalable quantum information technologies such as multi-node quantum networks[1-3], fault-tolerant quantum computing[4-6], and nanoscale nuclear magnetic resonance (NMR)[7-10]. One of the key challenges in enabling quantum applications is the ability to preserve the NV center's spin coherence in a noisy environment, as the gate fidelity is significantly limited by the decoherence process[4, 11-13]. Sophisticated microwave (MW) pulse techniques have been invented to protect the spin coherence[14-16] by exploiting certain properties of the bath noise [4,11]. More importantly, it was shown that it is key to consider the quantum nature of the bath dynamics in the design of MW pulse sequences to go beyond the current limit of the NV-based quantum information technologies [12, 17].

One essential technique to protect and control the spin coherence is dynamical decoupling (DD)[18-22]. Standard DD methods include the Hahn-echo(HE)[23] and Carr-Purcell-Meiboom-Gill (CPMG) sequences[24-26], which dynamically decouple the spin qubit from its noisy bath to prolong the spin coherence. Furthermore, more complex sequences[6, 27-29], such as the Waugh-Huber-Haeberlen sequence[27], ε-CPMG[28], and dynamical decoupling with radiofrequency waves[4-7, 30], have been applied to the NV center to effectively remove dipolar interactions, correct pulse errors, and detect nuclear spins within specific frequency bands, respectively. It is important to note that in general the DD method is an essential tool in noise spectroscopy[19, 31, 32]. Using the DD technique with a different number of pulses amounts to applying a filter function in the frequency domain, suppressing specific frequency components of the noise power spectrum. Therefore, an accurate understanding of the DD signal is critical not only for the qubit's coherence protection but also for the exact modeling of the bath dynamics.

Due to its utmost importance in quantum applications, NV decoherence processes in diamond have been extensively investigated, revealing that the electron spin associated with nitrogen donors (P1 centers) is one of the dominant sources of the decoherence[33-37]. P1 centers are common impurities in type-I diamonds[9, 38, 39], and they are generated at various concentrations when creating high-density NV centers using nitrogen implantation[38, 40]. Several experimental advancements have been reported in the past decade in detecting and characterizing the precise nature of P1 baths, including the characterization of the local density of P1 centers[41,42], and individual P1 centers around a NV center[43]. Additionally, the coherent manipulation of P1 centers has been demonstrated under the HE and DD pulse sequences, suggesting the potential use of P1 centers as ancilla qubits and reporter sensors[44-48].

Previous theoretical calculations of the P1-induced NV decoherence have employed two main approaches: a semi-classical theory[33, 49, 50] and a quantum bath theory[51-54]. The semi-classical theory approximates the bath dynamics as an effective magnetic noise, as e.g. in the Ornstein-Uhlenbeck (O-U) noise model, where the noise process is characterized as Markovian and Gaussian[50, 55]. The O-U noise model has been widely and successfully adopted to analyze the overall features of the NV decoherence in a P1 bath[33, 56-59]. On the other hand, the quantum bath model describes the decoherence as originating from the entanglement between the qubit and its environment. In particular, the quantum bath theory[60-63] combined with a cluster-correlation expansion (CCE) method[55, 64-69] has successfully predicted important features of the NV decoherence process without any empirical parameters to describe the noisy environment in diamond. One of the key parameters that reflect both the static and dynamical properties of the bath dynamics is the stretched exponent $p$, which is derived from the coherence function $L(t) = e^{-\left(\frac{t}{T_{2,echo}}\right)^p}$, where $T_{2,echo}$ is the HE coherence time For the P1-driven NV decoherence, $p$ is measured to be between 1.0 and 2.0 in experiments[41, 58]. The CCE method[52-54, 70] estimated $p$ to be between 1.0 to 1.6, showing a good agreement with the experimental finding.

The quantum theory of the P1-driven NV decoherence, however, remains incomplete. There are inconsistencies between the values of $T_{2,echo}$ and $p$ reported by previous CCE studies[52-54, 70], which need to be clarified. The $T_{2,echo}$ results obtained in two previous CCE studies[52, 53] differ by 35% to each other across a wide range of P1 concentrations from 1 ppm to 100 ppm. This discrepancy likely stems from the adoption of different computational methods, including the level of spin correlation considered in the calculations, the P1 center model, and the ensemble averaging scheme (EAS). J.C. Marcks *et al.*,[53] built a P1 center model with an effective hyperfine interaction and showed that $T_{2,echo}$ is numerically converged by accounting for four-spin correlations in the CCE calculation. H. Park *et al.*,[52] employed a P1 center model that explicitly incorporates a nitrogen nuclear spin of the P1 center and used CCE calculations accounting up to pair-P1 correlation. Most recently, Schätzle *et al.*,[54] developed a partitioned CCE method (pCCE), demonstrating that the stretched exponent $p$ varies between 1.0 and 1.6 depending on whether the hyperfine coupling is included in the P1 center model. These authors also utilized an effective P1 center model and showed that the numerical convergence was achieved at the pair-spin correlation level of the CCE theory. We also note that the inconsistencies between different theoretical results may arise from different fitting approaches adopted for the coherence function[70].

In addition, the P1 bath-driven NV decoherence dynamics under various DD sequences remains elusive. Notably, a semi-classical theory[50] predicted that the coherence time ($T_{2,n}$) as a function of the pulse number ($n$) in the CPMG DD scales as $T_{2,n} = n^\lambda T'_{2,echo}$, with the scaling exponent being $\lambda = 2/3$, regardless of the P1 concentration. However, this result is apparently inconsistent with previous experimental findings which reported different $\lambda$ values, between 0.4 and 0.7 [33, 56, 57, 71-73] for different P1 concentrations. In addition, the $\lambda$ values[71] shows a dependence on the P1 concentration. So far, no quantum bath theory calculations have been performed in the literature to address this discrepancy between theory and experiments. Thus, for a complete understanding of the NV decoherence in the P1 bath, it is necessary to establish how to use and converge the CCE method for the simulation of Hahn-echo signal and to extend the method to more complex DD sequences.

In this study, we perform a combined theoretical and experimental study to investigate the P1-induced NV spin ensemble decoherence under dynamical decoupling. Specifically, we compute the HE and CPMG signals of the NV centers by using the CCE method. For the HE signals, we confirm that the dominant mechanism determining the NV decoherence is pair-P1 correlation. We also find that the presence of a phase factor in a single-sample coherence function plays an important role in the ensemble NV decoherence. For the CPMG signal, we compute the $T_{2,n}$ and $p$ as a function of the number of pulses at different CCE levels of theory from CCE2 to CCE6. We find that NV decoherence is primarily governed by pair correlations at P1 densities lower than 100 ppm and when a number of pulses smaller than 12 is used; multiple-spin correlations become significant as the P1 density increases, and the pulse number grows. Importantly, we show that the scaling exponent $\lambda$ is not a constant but varies as a function of the pulse number $n$. Finally, we conduct CPMG measurements for two diamond samples with different P1 densities of 0.8ppm, and 13ppm. We find that the scaling exponent $\lambda$ increases with the number of pulses, showing a trend that is in excellent agreement with our theoretical predictions. Our findings uncover the significance of the quantum nature of the P1 bath dynamics in understanding the NV spin decoherence under the DD sequence. In addition, we provide a detailed understanding of the P1-bath-driven NV decoherence, laying a critical foundation for developing a complete noise model that could be used to optimize the NV performance in quantum applications.

## RESULT

**System and theoretical model.**

We adopt a central spin model to compute the NV decoherence time[55] in the presence of P1 centers. The NV center has an electron spin number S=1. The qubit state is defined as $|\psi\rangle = \frac{1}{\sqrt{2}}(|0\rangle + |-1\rangle)$, where $|\pm 1\rangle$ are the NV spin states whose degeneracy is lifted in a moderate magnetic field applied along the NV center's principal axis, which aligns with the [111] crystallographic direction. The P1 centers provide bath spins: it has an electron spin ($S = 1/2$) and a nuclear spin ($I = 1$ for $^{14}N$), which are coupled to each other by the hyperfine interaction. Notably, a P1 center in the diamond lattice undergoes a Jahn-teller (JT) distortion in one of four possible directions, which induces hyperfine anisotropy.

The coherence function $L(T)$ is defined as $L(T) = \langle \mathcal{J}|(U^{(-1)})^\dagger U^{(0)}|\mathcal{J}\rangle$, where $|\mathcal{J}\rangle$ represents a product state of the bath spins. Thus, $L(T)$ describes the overlap of the bath states bifurcated in their evolution, depending on the NV sublevels $|0\rangle$ and $|-1\rangle$ [55, 66]. The bath evolution operator $U^{(0/-1)}$ is determined by the applied MW pulse sequences. An $n$-pulse CPMG sequence applies a $\pi$ pulse to the central spin at each $t_i = \frac{(2i-1)T}{2n}$, where $i = 1,2,\ldots,n$, and T is the total free evolution time. A CPMG pulse with $n = 1$ simply corresponds to the Hahn-echo. Thus, the general bath evolution operator for the $n$-pulse CPMG sequence is expressed as $U^{(0/-1)} = \cdots e^{-iH^{(0/-1)}(t_3-t_2)}e^{-iH^{(-1/0)}(t_2-t_1)}e^{-iH^{(0/-1)}t_1}$, in which the bath Hamiltonian switches between $H^{(0)}$ and $H^{(-1)}$ at each time $t_i$ as the $\pi$ pulse flips the qubit states. To predict the decoherence dynamics of the many-body spin system, we employ the CCE method, which factorizes $L(T)$ into a product of cluster-correlation terms. The CCE method has been successfully applied to study various solid-state spin qubits[67, 74-77]. Details on the bath Hamiltonian, the CCE theory, and numerical convergence tests can be found in Methods and Supplementary Note 1, 2.

**CCE calculations for different P1 bath models.**

We start our investigation by comparing the NV decoherence obtained by different CCE methods and P1 models. We first analyze the impact of a phase factor, representing the precession of central spin[78] in a single-sample coherence function (see Supplementary Note 1.2 for details), on the ensemble average of the coherence functions. Then, we compare the different P1 bath models, which differ by how the nitrogen nuclear spin is treated (see Supplementary Note 1.3 for details). Finally, we address the numerical convergence of the CCE method for the calculation of the Hahn-echo signal.

Figure 1(a) shows $T_{2,echo}$ as a function of the P1 density from 1ppm to 100ppm, computed by using two different EASs: the average of the single-sample coherence functions is performed with and without the inclusion of the phase factor. Note that the EAS with and without the phase factor was adopted, respectively, in the previous study by Marcks *et al.*[53] and in our previous study[52]. At [P1] = 1ppm, we found $T_{2,echo}$ to be 286.31 $\mu s$ when including the phase factor, which is approximately 0.7 times smaller than that of the EAS obtained by neglecting the phase term. This ~0.7 times deviation is consistently observed across all P1 densities. We also find that the $T_{2,echo}$ results obtained by incorporating the phase factor are in better agreement with experiments. We note that the slopes of $T_{2,echo}$ on a logarithmic scale computed by the two EASs are almost the same (-1.04 and -1.06 with and without the phase term, respectively). Our results show that the phase information of the single-sample coherence function plays an important role in determining the decoherence time and its inclusion in the ensemble average is necessary to converge the CCE calculation.

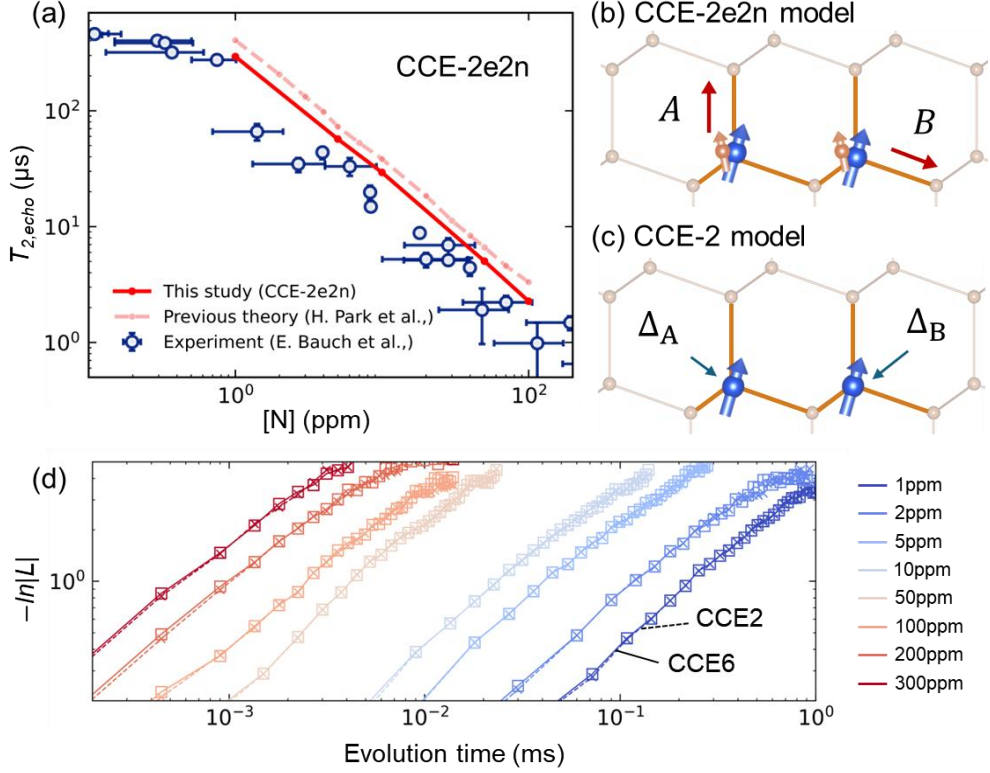

**Figure 1. Coherence times and P1 center models. (a)** Hahn-Echo coherence times ($T_{2,\ echo}$) calculated with two different EASs at the CCE-2e2n level of theory as a function of nitrogen concentration [N]. The red solid line is the result obtained by using the EAS with the phase factor, while the dashed red line shows our previous results[52], which omits the phase factor. Blue dots correspond to previous experimental results[58]. **(b, c)** Illustration of different P1 center models. Blue and orange arrows represent an electron spin and a nuclear spin, respectively. The P1 model in (b) includes both electron and nuclear spins as quantum spins. A and B denote the JT distortion direction, where A is on-axis and B is one of the three off-axes. In the P1 model in (c), the hyperfine interaction is included as an effective field for the electron spin while the hyperfine levels are not considered. $\mathbf{\Delta}_A$ and $\mathbf{\Delta}_B$ represents the effective fields for P1 centers distorted in A and B JT directions (See Supplementary Note 1.3). **(d)** NV coherence functions $L$ over a wide range of P1 concentration, [P1], computed by employing the CCE-2 (dotted line with x's) and CCE-6 (solid line with squares) theories.

**Table I. Summary of coherence time $T_{2,echo}$ and exponent $p$ (see text) computed with different CCE theories for P1 densities of 1, 10 and 100ppm.**

| CCE method | 1 (ppm) | | 10 (ppm) | | 100 (ppm) | |
|---|---|---|---|---|---|---|
| | $T_{2,echo}$ ($\mu s$) | $p$ | $T_{2,echo}$ ($\mu s$) | $p$ | $T_{2,echo}$ ($\mu s$) | $p$ |
| CCE-2e2n | 286.31 | 0.92 | 29.14 | 0.93 | 2.36 | 0.88 |
| CCE-2 | 236.33 | 0.99 | 24.43 | 0.95 | 2.51 | 0.94 |
| CCE-4 | 240.46 | 0.90 | 26.43 | 0.95 | 2.59 | 0.91 |
| CCE-6 | 236.20 | 0.98 | 24.15 | 0.94 | 2.47 | 0.93 |
| CCE-4 (Ref [53]) | 245.83 | 1.29 | 25.16 | 1.24 | 2.09 | 1.27 |
| pCCE(2,4) (Ref [54]) | ~256 | 1.04 | - | - | - | - |

Table I shows the $T_{2,echo}$ and $p$ results computed by considering two different P1 models, a *full* P1 model, including both electron and nuclear spins as quantum spins (see Figure 1b) and an *effective* P1 model, where the hyperfine interaction is included as an effective field for the electron spin, (Figure 1c) for P1 densities of 1ppm, 10ppm, and 100ppm. The CCE calculation includes pair-P1 correlations and the different P1 models are termed as CCE-2e2n and CCE-2 for the *full* P1 and *effective* P1 models, respectively. We find that both models make consistent predictions on the change of $T_2$ as a function of the P1 density: the slopes of $T_{2,echo}([P1])$ on the log scale are -1.00 and -1.04 for CCE-2 and CCE-2e2n, respectively, and the $p$ values are also consistent with each other. We observe, however, that the values of the $T_{2,echo}$ obtained in the two models differ as the P1 density decreases below 10 ppm. The differences in $T_{2,echo}$ between the models are 17.5%, 16.2%, and 6.4% for P1 densities of 1ppm, 10ppm, and 100ppm, respectively. In the full P1 model, the presence of nuclear spins induces hyperfine splitting of the energy levels. Our result shows that the P1's electron spin flip-flop transitions, the dominant source of the NV decoherence, are more suppressed in the presence of hyperfine splitting of the levels, compared to those obtained with the effective P1 model. This effect becomes significant when the dipolar interaction between the P1 centers is weakened as the P1 density decreases below 10 ppm. Our results suggest that the effective P1 model captures the essential, qualitative physics of the P1-driven decoherence with reasonable computational efficiency; however, more accurate predictions require incorporating hyperfine levels in the P1 model.

To identify the role of high-order spin-spin correlation in the Hahn-echo decoherence process, we perform CCE calculations from CCE-2 to CCE-6 levels of theory by employing the effective P1 model. Figure 1d directly compares the coherence functions computed at the CCE-2 and CCE-6 levels of theory. In Table I, we also present the $T_{2,echo}$ and $p$ values obtained with CCE-2, CCE-4, and CCE-6 theories at 1, 10, 100ppm Figure 1d shows that the CCE-2 and CCE-6 theories yield similar results over a wide range of [P1] from 1ppm to 300 ppm. In fact, the difference in coherence function between the two theories is negligible for all P1 densities throughout the entire evolution time. In Table 1, the computed $T_{2,echo}$ values exhibit only small fluctuations for increasing CCE orders, with differences within 10%, and the $p$ values remain consistently around ~0.9 across all CCE orders. The $T_{2,echo}$ values are also in a good agreement with previous results obtained by the pCCE(2,4)[54] and CCE-4[53] methods. However, we find that the computed $p$ values in Table 1 are slightly different from previous results. This minor discrepancies may be attributed to differences in the fitting approach, as highlighted in recent studies[70]. Overall, our results show that the dominant contribution to the HE decoherence is pairwise spin-spin interactions in the P1 bath and that the coherence function converges with the second-order CCE method.

**Coherence dynamics under multiple-pulse sequences**

We now turn to the P1 bath-induced NV decoherence under the CPMG sequence. We first investigate the dependence of the CPMG coherence time ($T_{2,n}$) on the P1 center model, where $n$ is the number of pulses. In Supplementary Figure 9, we compare the $T_{2,n}$ times for $n$= 1, 4, 16 and 64 computed using CCE-2 and CCE-2e2n for the P1 densities of 1ppm, 10ppm, 50ppm and 100ppm. For all P1 center densities, both P1 models exhibit the same trend in the scaling of $T_{2,n}$ with respect to the pulse number. This result indicates that the effective P1 center model remains robust even as the number of pulses increases across different P1 densities. Thus, we adopt the effective P1 model for the rest of this study.

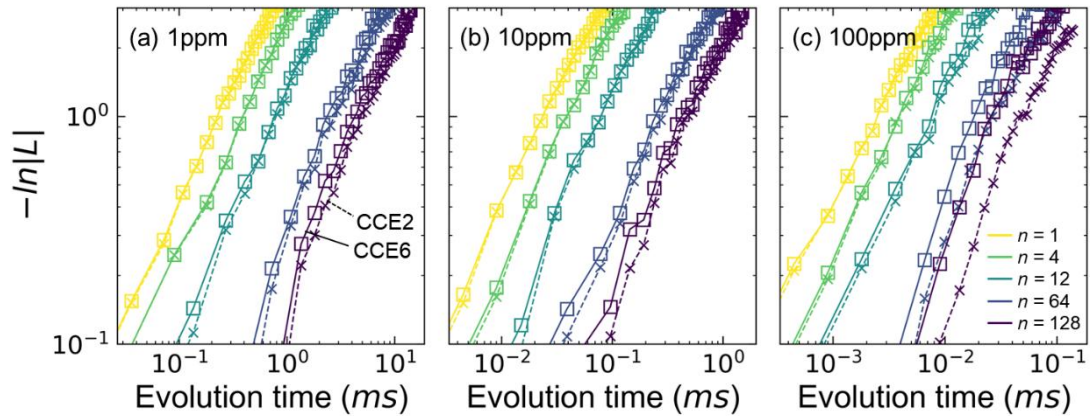

**Figure 2 Coherence of an NV spin ensemble under multi-pulse CPMG.** In the presence of (a) 1ppm, (b) 50ppm, and (c) 100ppm P1 densities, coherence functions are simulated by adopting CCE-2 (dotted line with x's) and CCE-6 theory (solid line with squares) for various numbers of pulses.

Figure 2(a) compares the coherence functions of an NV spin ensemble as obtained from the CCE-2 and CCE-6 theories under the CPMG sequence with $n$ =1,4,12,64 and 128, for the 1ppm P1 density. We find that the high-order spin correlation of P1 centers has a greater contribution to the NV decoherence when the number of pulses is large. For n = 1 or 4, we observe a negligible difference between the CCE-6 and CCE-2 results. We find, however, that the coherence predicted by the CCE-6 theory decays faster than that of the CCE-2 theory as the number of pulses increase to 12, resulting in a 4% difference between the CCE-6 and CCE-2 computed $T_{2,12}$ times, which are 776.8 $\mu s$ and 806.0 $\mu s$, respectively. With $n = 128$, the deviation between the two CCE theories grows, giving rise to a 16% difference between the respective $T_{2,128}$ values: 4452.1 $\mu s$ in CCE-6 and 5165.9 $\mu s$ in CCE-2. Our results highlight that the pairwise spin flip-flop transitions dominate the spin decoherence for *n* below 12, but the effect of high-order spin-cluster correlations become significantly important for *n* > 12, accelerating the NV coherence decay on top of the pair-spin contribution.

More remarkably, our calculations reveal that the deviation between the CCE-2 and CCE-6 results increases significantly at P1 densities beyond 10 ppm. As shown in Figure 2(b) and (c), under a 128-pulse CPMG sequence, the $T_{2,128}$ times are 513.66 $\mu s$ and 455.35 $\mu s$ for the CCE-2 and CCE-6 levels of theory at 10 ppm, and 51.48 $\mu s$ and 28.77 $\mu s$, respectively, at 100 ppm. A comparison of the coherence functions for additional P1 densities can be found in Supplementary Figure 10, where we observe the same trend. P1 spin baths may be characterized by the average dipolar interaction strength between the P1 centers within the bath. We find that the average interaction strength in P1 spin baths with different P1 densities ranges from 1.46 kHz at 1 ppm to 437.32 kHz at 300 ppm. Our results show that high-order spin-spin correlations become significant when the average dipolar interaction strength exceeds 14.58 kHz, corresponding to a P1 density of 10 ppm.

Figure 3a shows the computed $T_{2,n}$ times as a function of the pulse number ($n$) from 1 to 128 across a wide range of P1 densities (1 to 300ppm) on a double-logarithmic scale. We note that $T_{2,n=1}$ corresponds to $T_{2,echo}$ as shown in Figure 1a. For each P1 density, we observe that $T_{2,n}$ increases with *n* as the NV centers become further decoupled from the P1 bath under the CPMG sequence. Notably, however, the scaling behavior of $T_{2,n}$ with respect to *n* differs for various P1 densities. To characterize the scaling behavior, we first apply the conventional scaling law predicted by semi-classical theory[50, 56]: $T_{2,n} = n^\lambda T'_{2,echo}$. We then extract the $T'_{2,echo}$ parameter and the scaling exponent $\lambda$, as shown in Figure 3b. For P1 densities $[P1] \leq 50$ ppm, the $\lambda$ values remain nearly constant with an average $\bar\lambda$ of $0.72 \pm 0.02$. However, as $[P1]$ increases beyond 100 ppm, the $\lambda$ values decrease with an average $\bar\lambda = 0.62 \pm 0.05$. It is also worth noting that across all the P1 densities, the $\lambda$ values range from 0.56 to 0.78, consistent with previously reported experimental results[36-38, 48-50].

We emphasize, however, that the computed $T_{2,n}$ times shown in Figure 3a are not well described by a linear scaling law with a single scaling exponent. Instead, $T_{2,n}$ exhibits strong quadratic behavior as a function of *n* across all P1 concentrations. The non-linear behavior of $T_{2,n}$ becomes more evident when directly compared to the coherence time computed using linear scaling with $\lambda=2/3$[50, 56], as shown in Figure 3a. To better quantify the scaling, we compute the slope of $T_{2,n}$ on a logarithmic scale at each *n*, defined as: $Slope = \frac{\Delta(\log_{10} T_{2,n})}{\Delta(\log_{10} n)}$. This slope can be interpreted as a local $\lambda$, as a function of *n*. The results are displayed in Figure 3c and 3d for P1 densities below 100 ppm and above 100 ppm, respectively. Interestingly, for P1 densities below 100 ppm, the slope around *n* = 10 is similar to the semiclassical prediction of $\lambda=2/3$. However, for *n* > 10, the slope exceeds 2/3, implying that the decoupling effect predicted by the quantum theory is stronger than that of the semi-classical theory. In contrast, for P1 densities above 100 ppm, the decoupling effect is different: while the slope increases with *n*, it eventually saturates at $\lambda=2/3$ for *n* > 10, aligning with the results of the semiclassical theory.

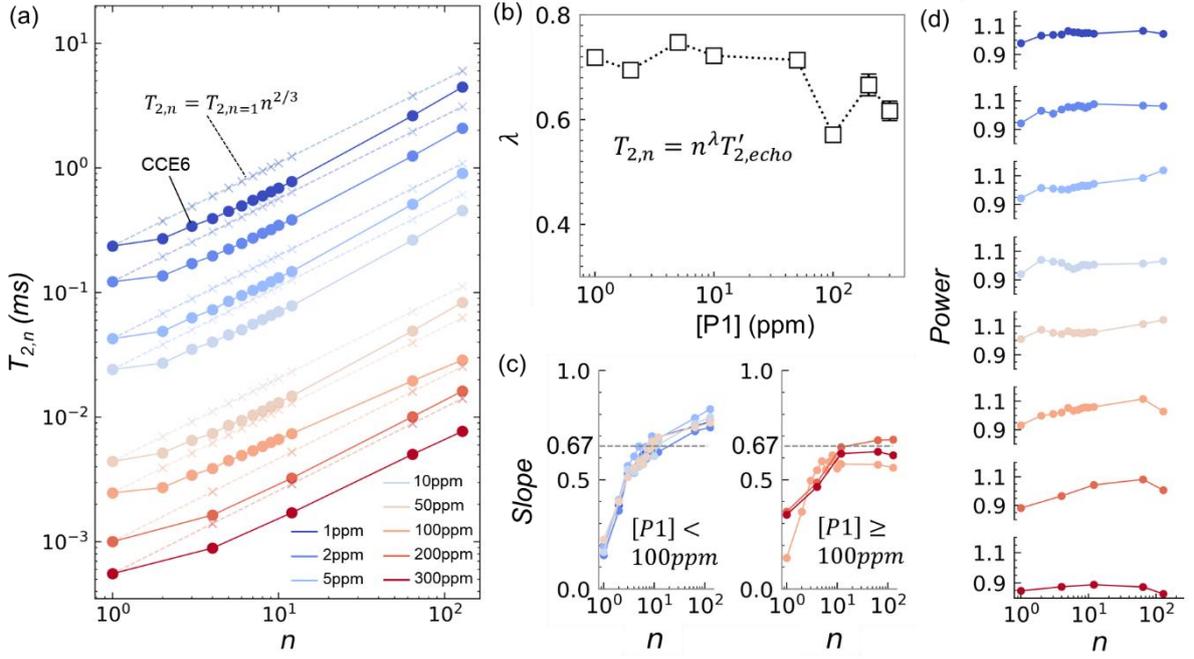

**Figure 3. CPMG coherence times of an NV spin ensemble** (a) Coherence times $T_{2,n}$ as a function of the number of pulses $n$ on a double-logarithmic scale for various P1 densities ranging from 1ppm (blue line) to 300ppm (red line). The coherence function is computed at the CCE6 level of theory and $T_{2,n}$ is obtained by fitting the coherence function to $L(T) = e^{(T/T_{2,n})^p}$. The solid line with circles indicates the predictions of our calculations, while the dotted line with crosses illustrates the coherence times predicted by the semi-classical scaling equation $T_{2,n} = n^{2/3} T'_{2,echo}$, where $T'_{2,echo} = T_{2,n=1}$. (b) $\lambda$ values as a function of the P1 density. $\lambda$ is obtained by fitting the function $T_{2,n} = n^{\lambda} T'_{2,echo}$. (c) Slopes of $T_{2,n}$ as a function of $n$, $Slope = \frac{\Delta(\log_{10} T_{2,n})}{\Delta(\log_{10} n)}$. The results for [P1] = 1 to 50 ppm and those for [P1] = 100 to 300 ppm are shown in the left and right panel, respectively. The gray dashed line represents a constant slope of 2/3. (d) The stretched exponents $p$ (Power) as a function of $n$ for P1 densities from 1ppm to 300ppm.

Figure 3d shows the stretched exponent $p$ as a function of the pulse number for different P1 densities. For P1 densities ranging from 1 ppm to 50 ppm, the $p$ value increase with $n$: $p$ values are $0.96 \pm 0.03$, $1.05 \pm 0.03$, and $1.08 \pm 0.05$ at $n = 1$, $n = 12$, and $n = 128$, respectively. This increase in $p$ reflects an enhanced coherence protection by dynamical decoupling over short timescales. However, for P1 densities above 100 ppm, the $p$ value reaches a maximum around n = 64: the $p$ values are $0.89 \pm 0.04$, $1.00 \pm 0.09$, and $0.95 \pm 0.11$ at $n = 1$, $n = 12$, and $n = 128$, respectively. Notably, for a P1 density of 300 ppm, $p$ values remain below 0.9 across all pulse numbers. Together with the analysis of the local slope of $T_{2,n}$ shown in Fig. 3c, these results indicate that the decoupling effect undergoes a qualitative change when the P1 density exceeds 100 ppm.

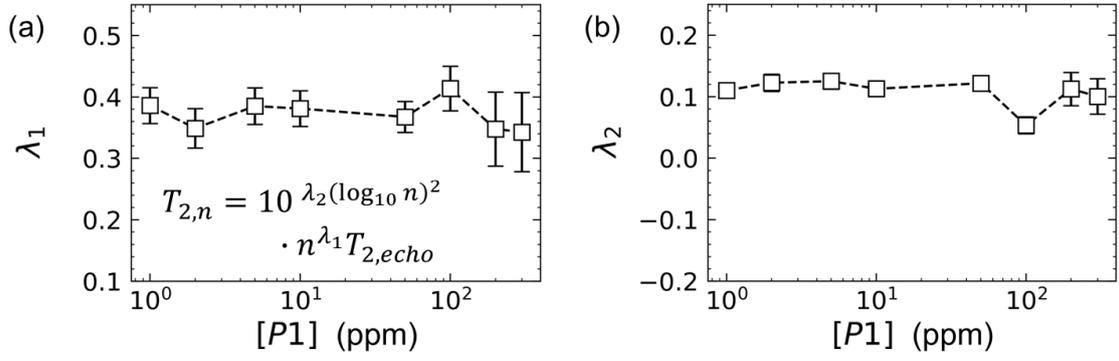

**Figure 4. Scaling exponents for the coherence time under Dynamical Decoupling.** $\lambda_1$ (a) and $\lambda_2$ (b) are the scaling exponents used to fit the data obtained from equation (2), and are shown as a function of the P1 concentration [P1].

While the previous analysis reveals the P1-density-dependent decoherence behavior, it is useful to generalize the scaling law described above, to characterize the variation of $T_{2,n}$ more precisely. In Eq. 1, we introduce a new scaling equation with two independent parameters $\lambda_1$ and $\lambda_2$ to capture the quadratic behavior of $T_{2,n}$:

$$\log_{10} T_{2,n} = \log_{10} T_{2,echo} + \lambda_1 \log_{10} n + \lambda_2 (\log_{10} n)^2 \tag{1}$$

$$T_{2,n} = T_{2,echo} n^{\lambda_1} \cdot 10^{\lambda_2 (\log_{10} n)^2} \tag{2}$$

Figure 4a and 4b present the computed $\lambda_1$, and $\lambda_2$ values. We find that $\lambda_1$, and $\lambda_2$ are nearly constant across P1 densities ranging from 1 ppm to 300 ppm, with average values of $\lambda_1 =$ $0.37 \pm 0.02$ and $\lambda_2 = 0.11 \pm 0.02$. Notably, the finite $\lambda_2$ value confirms the quadradic dependence of $T_{2,n}$.

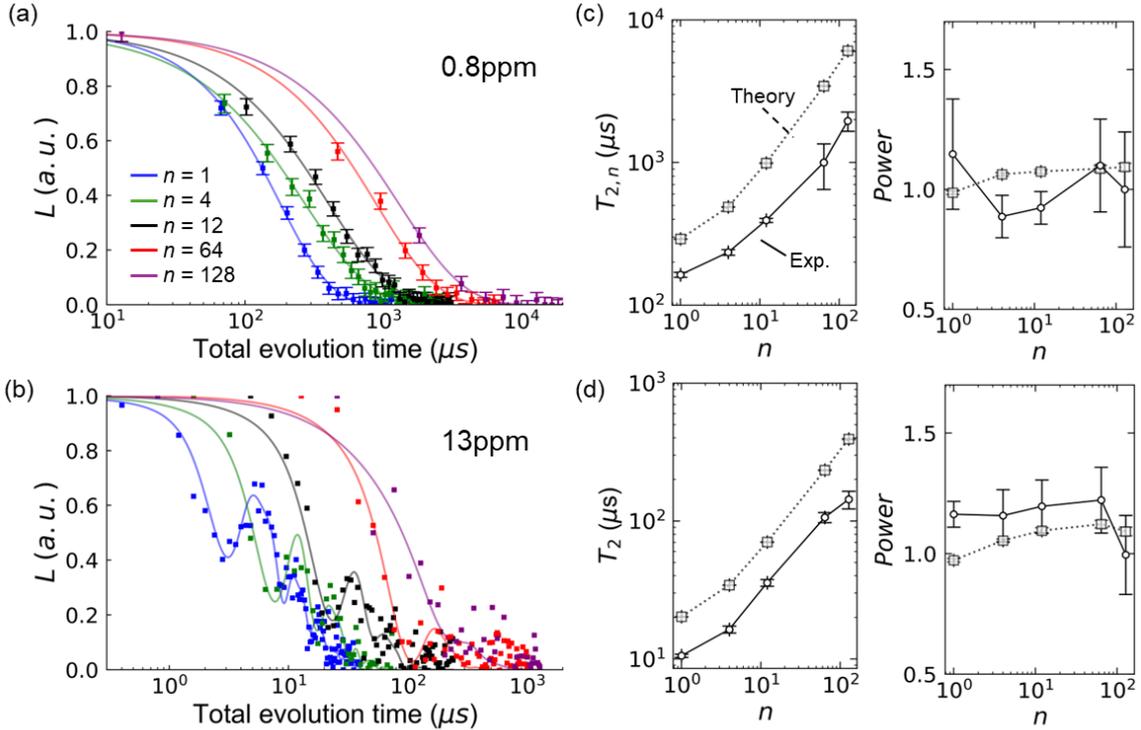

**Figure 5 Comparison of experimental and theoretical results. (a, b)** Optically measured coherence functions for diamond sample with [P1] = 0.8ppm and [P1] = 13ppm for various number of pulses. Square dots and solid lines indicate experimental data and fitting results, respectively. Each color—blue, green, black, red, and purple—corresponds to specific pulse numbers: 1, 4, 12, 64, and 128, respectively. (b, d) Comparison of coherence times ($T_2$) and stretched exponent (Power) obtained experimentally (solid line with circles) with our CCE-6 theoretical predictions (dotted line with squares) as a function of the pulse number n for [P1] = 0.8ppm (c) and for 13ppm (d).

**Optically measured NV spin coherence under dynamical decoupling technique**

To validate our theoretical findings, we conduct experiments on two diamond samples. Figures 5a and 5b show the measured NV coherence functions for samples with P1 concentrations of 0.8ppm and 13 ppm, respectively, under various pulse numbers ($n = $ 1, 4, 12, 64, and 128). The measured $L(t)$ reveals that the envelope decay of the NV coherence is suppressed as the pulse number increases. The spin coherence envelope follows a stretched exponential function with electron spin echo envelop modulations (ESEEM):

$$L(t) = \exp\left(-\left(\frac{t}{T_{2,n}}\right)^p\right)\left(1 - b\sin^2\left(\frac{\omega_1 t}{2}\right)\sin^2\left(\frac{\omega_2 t}{2}\right)\right)$$

, where $\omega_1$ and $\omega_2$ are ESEEM frequencies due to the coupling between the central NV spin and the nearby $^{13}C$ nuclear spins; $b$ is the modulation depth. The first term with two parameters ($T_{2,n}$ and $p$) describes the coherence decay caused by the P1 bath, which is the dominant noise source in our diamond samples, in contrast to the slower dynamics of the nuclear spin bath.

Figures 5c and 5d present the $T_{2,n}$ and $p$ values extracted from the experimental data and those computed using our theoretical models for the 0.8 ppm and 13 ppm samples, respectively, plotted as a function of the number of pulses. For the Hahn-echo ($n=1$) case, the experimentally measured coherence times are $T_{2,n=1}^{(exp)}$ =163.3 $\mu s$ for the 0.8 ppm sample and 10.5 $\mu s$ for the 13 ppm sample. As $n$ increases, $T_{2,n}^{(exp)}$ initially rises gradually, then grows more rapidly, reaching $T_{2,n=128}^{(exp)} = 1955.0$ $\mu s$ and $T_{2,n=128}^{(exp)} = 143.3$ $\mu s$ for 0.8 ppm and 13 ppm samples, respectively. For $n = 1$, our theoretical results are 291.3 $\mu s$ and 20.2 $\mu s$, differing by 10.2% and 21.7% from the experimental results on logarithmic scale for 0.8 ppm and 13 ppm, respectively. This discrepancy between theory and experiment remains consistent for larger $n$ within an uncertainty of 5% and 7% for 0.8 ppm and 13 ppm, respectively. The observed deviations between theoretical and experimental $T_{2,n}$ are likely due to additional paramagnetic defects in the diamond samples that can be created during sample synthesis processes, which are not accounted for in our theoretical model[52].

Importantly, we find that the variation of $T_{2,n}$ observed in the experimental data follows the non-linear behavior predicted by our theoretical framework, as shown in Fig. 5b and 5d. To characterize this behavior, we compute the scaling exponents $\lambda_1$ and $\lambda_2$ using Eq. (2) for both theoretical and experimental data. For the 0.8 ppm diamond sample, the experimental scaling exponents are $\lambda_{1,exp} = 0.16 \pm 0.22$ and $\lambda_{2,exp} = 0.16 \pm 0.1$. Similarly, for the 13 ppm sample, we found $\lambda_{1,exp} = 0.41 \pm 0.14$ and $\lambda_{2,exp} = 0.08 \pm 0.06$. Theoretical predictions yield $\lambda_1 = 0.34 \pm 0.05$, $\lambda_2 = 0.14 \pm 0.02$ for the 0.8 ppm sample, and $\lambda_1 = 0.37 \pm 0.06$, $\lambda_2 = 0.12 \pm 0.03$ for the 13 ppm sample. The close agreement between experimental and theoretical scaling exponents confirm the quadratic scaling behavior of dynamically decoupled NV coherence in the P1 bath.

## CONCLUSION

In summary, we presented a combined theoretical and experimental investigation of the decoherence dynamics of an NV spin ensemble in the presence of P1 centers across a wide range of P1 densities from 1 ppm to 300 ppm. For the Hahn-echo pulse sequence, we examined differences in CCE theories and P1 center models to explain the coherence time discrepancies reported in previous studies. Our findings provide a unified picture, showing that P1 pair-spin correlations and the inclusion of the phase factor in the EAS scheme are crucial to accurately predict the NV decoherence under the Hahn-echo sequence. We also demonstrated that the effective P1 center model provides an efficient and reasonably accurate approach for predicting the P1-bath-driven NV decoherence dynamics across various pulse numbers, compared to the full P1 center model.

For the CPMG signal with $n \geq 1$, we investigated the NV decoherence for pulse numbers ranging from 1 to 128. Our findings revealed that sixth-order spin correlations are essential for accurately predicting the decoherence dynamics of NV spin qubits, specifically, in the case when the pulse number exceeds 12 and $[P1] \geq 100\ ppm$. Most importantly, our quantum bath theory predicted a non-linear scaling behavior of $T_{2,n}$, which was confirmed by our experimental measurements. Such behavior indicates that the scaling exponent $\lambda$ is a function of $n$ rather than a constant, suggesting that the P1 bath dynamics evolves differently for different pulse numbers. This observation aligns with the understanding that bath dynamics are indirectly influenced by the MW pulses applied to the NV center through its interactions with the P1 bath. Notably, the non-linear scaling behavior of $T_{2,n}$ is not captured by the semi-classical theory, demonstrating the significance of the quantum nature in the NV decoherence under dynamical decoupling.

In conclusion, our theoretical framework successfully elucidates the decoherence dynamics of NV spin ensembles under a P1 bath. Building on these findings, future work could explore the effects of other paramagnetic defects in diamond samples, in addition to the P1 bath, as well as investigate newly designed pulse sequences for various applications. These extensions are expected to enable more sophisticated noise suppression strategies and further optimization of the NV performance in quantum technologies.

## Method

**Experimental details**

The diamond samples used in the CPMG measurements are quantum-grade single-crystal, CVD in-grown NV ensemble samples (1.1% $^{13}$C) with [P1] = 13 ppm, [NV] = 4.5 ppm (DNVB14) and [P1] = 0.8 ppm, [NV] = 0.3 ppm (DNVB1) from Element 6.

We used a home-built confocal microscope setup the NV ensemble sample CPMG measurements. Both samples are illuminated with 532nm laser (CNI MGL-FN-532) pulses for initialization and readout generated by double-pass AOM (G&H 3200-121). All measurements were performed at room temperature under a magnetic field of 295G for the [P1] = 13ppm sample and 284.4G for the [P1] = 0.8ppm sample, both aligned parallel to the z-axis of the NV center with a misalignment angle of less than 0.5 degrees. These conditions closely match those used in the theoretical calculations (B = 300G). We apply microwave (MW) signals to NV ensemble electron spin manipulation between two states $m_s$ = 0 and $m_s$ = -1 by using the vector signal generator (Rohde & Schwarz SMBV100A) with I/Q modulation for the phase control. MW signal was pulsed with switch (ZASWA-2-50DR+) and amplified with 16W amplifier (ZHL-16W-43s+). Finally, NV spin state dependent fluorescence signal was detected by the avalanche photon detector (SPCM-AQRH). Whole experimental setup was controlled by the NI FPGA (NI-6583).

We performed CPMG-n measurement by repeating blocks of $\left(\tau - (\pi)_y - 2\tau - (\pi)_y - \tau\right)^{\times \frac{n}{2}}$ sequences, enclosed by two $\left(\frac{\pi}{2}\right)_x$ pulses. For [P1] = 13ppm, DNVB14 sample, we measured full NV spin coherence decay curve with ESEEM (Electron Spin Echo Envelope Modulation) from the bath of 13C nuclear spins and extracted $T_{2,n}$ by fitting it to the stretched exponential function with the ESEEM. For [P1] = 0.8ppm, DNVB1 sample, due to its long $T_{2,n}$ time compared to the ESEEM period, we detected coherences only at the maximum revival peaks, which occurs at $\tau = \frac{(2k-1)\pi}{2\omega_L}$, where k is the revival peak position index and $\omega_L$ is the 13C bath spin Larmor frequency, to isolate the decaying envelope. We then fitted data with the stretched exponential function on the envelope to extract $T_{2,n}$.

**Cluster-correlation expansion method with quantum bath model**

The decoherence dynamics of an NV spin ensemble is computed by utilizing the single-sample CCE method based on the quantum bath theory. In the quantum bath theory, the NV spin is prepared in a state $|\psi\rangle = \frac{1}{\sqrt{2}}(|0\rangle + |-1\rangle)$ after lifting the degenerate NV spin state $|\pm 1\rangle$ with a moderate magnetic field of 500G in the Hahn-echo sequence and 300G in the CPMG sequence. In the rotating frame with respect to the NV spin Hamiltonian, the total spin Hamiltonian includes the bath Hamiltonian and the interaction Hamiltonian between the qubit and the bath. The form of the total spin Hamiltonian depends on the model chosen for the P1 center. We consider P1 centers randomly distributed with equal probability around the NV center, assuming that the electron spin of a P1 center is fully localized at the atomic sites. To obtain the ensemble NV coherence function, we average the single NV coherence function calculated for a given spatial configuration over 20 different bath configurations. Additionally, we exploit the CCE method within a single-sample approach to account for the Ising-like coupling between the spins in a cluster and all other bath spins. Here, the initial P1 bath states (which we call a bath state template) are randomly sampled from a completely mixed thermal ensemble. A single NV coherence function with a given spatial configuration is obtained by averaging over 20 different bath state templates, all with the same spatial configuration. The convergence test of coherence function has been done for the number of spatial configuration and bath state templates (See Supplementary Note 2). Finally, the coherence time $T_{2,n}$ of an NV spin can be obtained by fitting the function $L(T) = e^{-\left(\frac{T}{T_{2,n}}\right)^p}$. In Supplementary Note 1, details of our CCE computation and quantum bath model of P1 bath are provided.

**Numerical convergence test for CCE theory**

The conventional convergence tests of CCE calculations are typically performed as a function of the bath radius ($r_{bath}$) and dipolar radius ($r_{dip}$), which define the cutoff for the P1 bath size and the maximum dipole-dipole interaction distance between two P1 centers, respectively. However, in simulations involving higher-order CCE-$k$ theories ($k > 2$), the number of clusters at the $k$-th order increases significantly, posing computational challenges. To address this computational challenge while preserving the essential P1 center clusters, we introduced an additional numerical parameter: the maximum number of clusters ($N_k$) at the CCE-$k$ expansion level.

Our convergence strategy is the following: after performing convergence tests for $r_{bath}$ and $r_{dip}$, we generate a connectivity map to determine whether the distance between any two P1 centers falls within $r_{dip}$. For a given $N_k$, $k$-th order clusters are generated based on this connectivity map. The $N_k$ cutoff is then applied to limit the number of clusters after sorting the generated clusters in descending order, according to their total dipolar coupling strength. Detailed results of the convergence tests for $N_k$ and other numerical parameters are provided in Supplementary Note 2.

For our CCE calculations, we developed a in-house software package called CCEX using C/C++ programming languages, EIGEN library, and MPI parallelization scheme. We also used UTHASH, a hash table library, for a fast clustering algorithm[79]. The source code is publicly available on our GitHub repository[80]. To cross-check the simulation results, we compare our Hahn-echo (HE) coherence results to those obtained using the gCCE method as implemented in PyCCE, previously developed by O. Mykyta et al.,[81]. The results are summarized in Supplementary Table I, and we found that the $T_{2,echo}$ and $p$ values for 1 ppm and 10 ppm P1 densities computed using the two codes are identical.

## Acknowledgements

This study is supported by the National Research Foundation (NRF) of Korea grant funded by the Korean government (MSIT) (No. 2023R1A2C1006270 and No. RS-2024-00442710), by Creation of the Quantum Information Science R&D Ecosystem (Grant No. 2022M3H3A106307411) through the NRF of Korea funded by the Korea government (MSIT). This material is based upon work supported in part by the KIST institutional program (Project. No. 2E33541). This work was supported by Institute of Information & communications Technology Planning & Evaluation (IITP) grant funded by the Korea government (MSIT) (No.2022-0-01026). This work was supported by the National Supercomputing Center with supercomputing resources including technical support (KSC-2024-CRE-0526), and Global-Learning & Academic research institution for Master's PhD students and Postdocs (G-LAMP) Program of the National Research Foundation of Korea (NRF) grant funded by the Ministry of Education (No.RS-2023-00285390). This research was supported by the education and training program of the Quantum Information Research Support Center, funded through the National research foundation of Korea (NRF) by the Ministry of science and ICT(MSIT) of the Korean government (No.2021M3H3A103657313). The work of MO and GG was supported by NSF QuBBE Quantum Leap Challenge Institute (Grant No. NSF OMA-2121044).